\documentclass{article}

\usepackage{arxiv}

\usepackage[utf8]{inputenc} 
\usepackage[T1]{fontenc}    
\usepackage[hidelinks]{hyperref}       
\usepackage{url}            
\usepackage{booktabs}       
\usepackage{amsfonts}       
\usepackage{nicefrac}       
\usepackage{microtype}      
\usepackage{lipsum}
\usepackage{graphicx}
\usepackage{natbib}
\usepackage{float}
\graphicspath{ {./images/} }

\usepackage{amsmath}
\usepackage[english]{babel}
\usepackage{array,tabularx}
\usepackage{makeidx}  
\usepackage{xparse}
\usepackage{moredefs}
\usepackage{lips}
\usepackage{epstopdf} 
\usepackage{color}
\usepackage{csquotes}
\usepackage{xcolor}
\usepackage{times}
\usepackage[nolist,nohyperlinks]{acronym}
\usepackage{comment}
\usepackage{paralist}
\usepackage{cleveref}
\usepackage{wrapfig}
\usepackage{multirow}
\usepackage{listings}
\usepackage[super]{nth}
\usepackage{algorithmic}
\usepackage{textcomp}

\crefformat{footnote}{#2\footnotemark[#1]#3}

\expandafter\def\expandafter\UrlBreaks\expandafter{\UrlBreaks
  \do\a\do\b\do\c\do\d\do\e\do\f\do\g\do\h\do\i\do\j%
  \do\k\do\l\do\m\do\n\do\o\do\p\do\q\do\r\do\s\do\t%
  \do\u\do\v\do\w\do\x\do\y\do\z\do\A\do\B\do\C\do\D%
  \do\E\do\F\do\G\do\H\do\I\do\J\do\K\do\L\do\M\do\N%
  \do\O\do\P\do\Q\do\R\do\S\do\T\do\U\do\V\do\W\do\X%
  \do\Y\do\Z}

\newcolumntype{L}[1]{>{\raggedright\arraybackslash}p{#1}}   
\newcolumntype{C}[1]{>{\centering\arraybackslash}p{#1}}     
\newcolumntype{R}[1]{>{\raggedleft\arraybackslash}p{#1}}    

\title{QC-Adviser: Quantum Hardware Recommendations for Solving Industrial Optimization Problems}

\author{
 Djamel Laps-Bouraba \\
  Databases and Information Systems\\
  University of Hagen\\
  Hagen, Germany \\
  \texttt{info@dbouraba.de} \\
   \And
 Markus Zajac \\
  Institute of Software Technology\\
  German Aerospace Center (DLR)\\
  Cologne, Germany \\
  \texttt{markus.zajac@dlr.de} \\
  \And
 Uta St{\"o}rl \\
  Databases and Information Systems\\
  University of Hagen\\
  Hagen, Germany \\
  \texttt{uta.stoerl@fernuni-hagen.de} \\
}

\begin{document}
\maketitle
\begin{abstract}
The availability of quantum hardware via the cloud offers opportunities for new approaches to computing optimization problems in an industrial environment. However, selecting the right quantum hardware is difficult for non-experts due to its technical characteristics. In this paper, we present the QC-Adviser prototype, which supports users in selecting suitable quantum annealer hardware without requiring quantum computing knowledge.
\end{abstract}


\section{Introduction}
Quantum computers have become increasingly important in recent years. Cloud providers that make quantum hardware available to everyone are increasingly attracting the interest of the industry. Companies and consortia are identifying potential applications for quantum computing in various sectors, such as logistics or material science~\citep{DBLP:journals/access/SinghDSPHCS24,bayerstadler2021industry}.

Today's quantum computers are in their infancy. They are limited in the number of qubits and produce errors in their calculations~\cite{businesses3040036,DBLP:journals/access/SinghDSPHCS24}. Therefore, ensuring a possible advantage over classical solutions is a challenge~\cite{businesses3040036}. This makes it difficult for companies to develop business models for investments in quantum technologies~\cite{businesses3040036}, and to demonstrate the necessary business impact of quantum technologies~\cite{bayerstadler2021industry}. 
In~\cite{DBLP:conf/gi/ZajacRS24}, possible approaches to identifying a potential advantage when using gate-based quantum computers are discussed.

However, companies and consortia expect quantum computing technologies could have a significant business impact. Especially because research is being carried out on improved hardware~\cite{DBLP:journals/access/SinghDSPHCS24}. The two expected effects are of particular interest to the industry:

\begin{compactenum}
    \item Solution quality and efficiency: Today's algorithms often calculate a local optimum, and therefore not an optimal solution~\cite{Tripathi2025}. Quantum technologies, such as quantum annealing, promise better solution quality and solutions for problems with large parameter ranges~\citep{bayerstadler2021industry}. The production and logistics environment is particularly affected by this, as better quality solutions contribute to process and cost efficiency.
    \item Faster solutions to optimization problems: Quantum computers can achieve a speed advantage over classical solutions when solving certain problems~\citep{DBLP:journals/access/SinghDSPHCS24,10.1007/978-3-031-37963-5_19}. In the case of quantum annealing, the aspect of acceleration has not been clarified conclusively and is being discussed~\citep{Yarkoni.2022}. The further development of improved quantum hardware could lead to enhanced devices in the future.
\end{compactenum}

Regardless of the question of an advantage, the question of how to select the best possible quantum hardware currently available must also be answered. 
In this paper, we focus on the selection of suitable quantum annealer hardware. This type of quantum hardware is often used to solve industrial optimization problems, as Yarkoni et al.~\cite{Yarkoni.2022} argue. The hardware itself is conveyed by so-called \textit{solvers}, which are hardware resources for problem solving. For business users who are not familiar with quantum technology, it is not easy to identify a solver. They have to deal with the following questions (we briefly discuss the technical terms in Section~\ref{preliminaries}):

\begin{compactenum}
    \item How many qubits are required?
    \item Can a hybrid or a QPU solver be used, or both?
    \item Is a decision accompanied by a benchmark score?
    \item How expensive is the use of a solver?
\end{compactenum}

To support users in the decision-making process, we have developed a software prototype that we call the QC-Adviser. This dialog-based tool requests problem-specific information, makes an estimate of the resources required, and presents the most appropriate solvers to the user. The contribution of our paper is the following:

\begin{compactenum}
    \item We describe our prototype for decision-making. This enables business users to identify suitable solvers for a range of problems.
    \item We show that users hardly needs any quantum computing knowledge to operate the QC-Adviser. The goal of the QC-Adviser is to hide as much specifics as possible by simple user guidance.
\end{compactenum}

The remainder of this paper is organized as follows: In Section~\ref{preliminaries}, we briefly cover the fundamentals. Related work is discussed in Section~\ref{related}. The general workflow of the QC-Adviser prototype is introduced in Section~\ref{workflow}. In Section~\ref{prototype}, we demonstrate its functioning using the TSP (Travelling Salesman Problem) and provide a brief summary in Section~\ref{conclusion}.
\section{Fundamentals}\label{preliminaries}
Quantum computers can be divided into gate-based quantum computers (circuit model) and quantum annealers (QA). QA rely on continuous-time evolution of a quantum system. Although gate-based quantum computers can handle a wider range of problems, QA are specifically designed to solve combinatorial optimization problems~\citep{Sehrawat_2024,Yarkoni.2022}. While classical computers are based on CPUs and bits, quantum computers are based on QPUs, which in turn contain a specific number of qubits. QA QPUs follow specific graph topologies, which in turn determine the layout and connectivity of the qubits~\citep{Yarkoni.2022}. Different topologies can lead to differences in the accuracy of the results for the same problem.

In order to calculate optimization problems on a QA, they must be formulated accordingly beforehand. There is already a proven approach to this: Optimization problems can be reduced to the so-called Quadratic Unconstrained Binary Optimization (QUBO) problems~\citep{Yarkoni.2022,DBLP:journals/anor/GloverKHD22}. It is an optimization formula that expresses a sum. The sum is based on binary variables and the products of pairs of these binary variables. Products are restricted to quadratic relationships between binary variables. In addition, each term has a coefficient.

We now turn to the TSP and its QUBO formulation as a representative example. The TSP can be described as follows: Given a set of nodes and edges (distances), the problem is to find the shortest possible tour $T$ that visits each node exactly once and returns to the starting node~\citep{Stogiannos2022}. A tour is described as:
\begin{equation}
T = (p_{1},...,p_{n},p_{n+1}), p_{n+1} = p_{1},
\end{equation}
\noindent
where $n$ is the number of nodes and $p_{i}$ is the node in the $i$th position of the tour. A binary variable is defined in the context of TSP as follows~\citep{Stogiannos2022}:
\begin{equation}
x_{v,p} = \left\{
\begin{array}{ll}
1 & \textrm{node $v$ is at position $p$ in the tour,} \\
0 & \, \textrm{otherwise.} \\
\end{array}
\right.
\end{equation}
The final QUBO formulation can be summarized as follows (detailed formulation can be found in the work of~Stogiannos et al.~\citep{Stogiannos2022}):
\begin{equation}\label{eq_total}
H_{TSP} = c_{1}(H_{v,p}+H_{p,v}+H_{E^{C}}) + c_{2}H_{W}
\end{equation}

The sum terms encode the following constraints and the minimization objective:

\begin{compactitem}
    \item $H_{v,p}$: Each node must appear at exactly one position on the tour.
    \item $H_{p,v}$: Each position of the tour must be occupied by exactly one node.
    \item $H_{E^{C}}$: The tour must consist of edges that really exist.
    \item $H_{W}$: Calculates the cost of a tour (minimization objective necessary to converge to the tour with the minimal cost).
\end{compactitem}

A problem formulated as a QUBO is then embedded in a QA QPU to be solved. For the overall process, from the definition of QUBO to the readout of the result, we refer to the work of~Yarkoni et al.~\cite{Yarkoni.2022}. The number of qubits required for embedding can be estimated. Stogiannos et al.~\cite{Stogiannos2022} present the estimation approach in the case of the TSP.

Based on an estimate, suitable solvers can be inferred. Thereby, QPU solvers represent different QPUs with a specific number of qubits and topologies. Hybrid solvers combine classical algorithms and QPUs. The most widely used solvers today come from the vendor \textit{D-Wave}~\citep{Yarkoni.2022, Tripathi2025}, which has several different solvers in its portfolio\footnote{\url{https://docs.dwavequantum.com/en/latest/industrial\_optimization/index\_get\_started.html\#opt-index-get-started}}.
\section{Related Work}\label{related}
There are several works on the question of how to best select the currently available quantum hardware for a given problem.

Salm et al.~\cite{DBLP:conf/closer/SalmBLW23}~deal with the question of how compilers and quantum computers can be automatically selected before the input circuit is compiled. To answer this question, the authors utilize machine learning methods to predict the accuracy of execution results on different quantum computers. The overall solution defines a process and an implemented software architecture based on the previous work of the authors. The \textit{NISQ Analyzer} is the central component of the architecture~\cite{DBLP:conf/closer/SalmBLW23,DBLP:conf/summersoc/SalmBBLWW20}. This component analyzes and selects a suitable implementation and quantum computer for a quantum algorithm chosen by the user~\cite{DBLP:conf/summersoc/SalmBBLWW20}. This work considers gate-based quantum computers and their inputs in the form of circuits. Quantum annealers and problems formulated in QUBO have not yet been taken into account. In our work, we offer support for this case.

Poggel et al.~\cite{DBLP:conf/qsw/PoggelQBWL23} examine the selection of the most suitable options (such as encoding, algorithm, and quantum hardware) that must be determined for solving optimization problems with quantum computers. To this end, they developed a framework that suggests various options as solution paths. Solution paths begin with the problem formulation and end with compilation and hardware selection. The actual selection of quantum hardware can be done by a component such as the \textit{MQT Predictor}~\cite{DBLP:conf/qsw/PoggelQBWL23,10.1145/3673241}. The \textit{MQT Predictor} uses a trained model (whose training data consists of quantum circuits from a wide range of applications) to select the hardware~\cite{10.1145/3673241}. The \textit{MQT Predictor} offers support for gate-based quantum computers. Although QUBO problem encoding is possible within the framework mentioned above, gate-based quantum hardware is selected. For this purpose, the problem is reformulated or converted as QAOA. Quantum annealers (such as those from \textit{D-Wave}) directly supports QUBO formulations. We take advantage of this fact in our work.

In general, benchmarks can also play a role in the selection of quantum hardware~\cite{DBLP:conf/sigsoft/WederBLSV20}. Several benchmarks have already been developed for quantum computing~\cite{lorenz2025systematicbenchmarkingquantumcomputers}. Looking at existing benchmarks for applications, it is striking that most were originally developed for gate-based systems and for specific problems. Some, such as \textit{Q-score}, have been extended for quantum annealing~\cite{lorenz2025systematicbenchmarkingquantumcomputers}. \textit{Q-score} is also used to determine the quality of the solution. We take the idea of solution quality and utilize it when selecting suitable solvers. The prerequisite is that this is known for a problem instance.

\section{Workflow Concept}\label{workflow}
\begin{figure}[H]
 \centering
 \includegraphics[width=\linewidth]{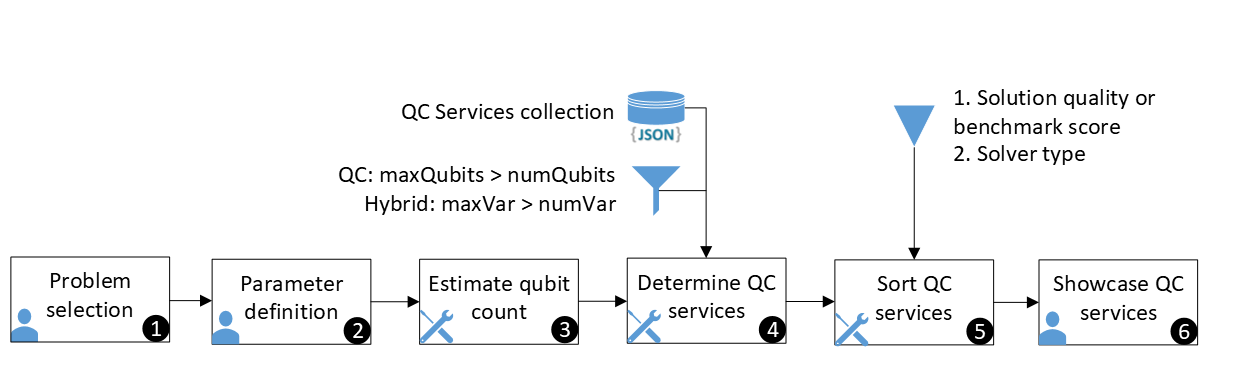}
 \caption{The workflow contains six steps. A person symbol stands for the user's interaction with the QC-Adviser, a tool symbol stands for the calculations performed by the QC-Adviser.}
 \label{fig:wfl}
\end{figure}

As mentioned in Section~\ref{preliminaries}, the number of qubits required can be estimated. Thereupon, possible solvers can be suggested to a user on a problem-specific basis. This fact forms the basis for the general workflow of the QC-Adviser prototype, which is shown in Figure~\ref{fig:wfl}. Detailed description of the individual steps follows.

\noindent
\textit{Step~1.} The user selects the problem to be solved and can read the description of the problem on request. Different problems are grouped into classes. The QC-Adviser currently offers the three problem classes \textit{Routing Problems}, \textit{Sequencing Problems} and \textit{General Problems}. General problems are the basic problems themselves, without any extension in the form of constraints, which only contain the implicit constraints. 

\noindent
\textit{Step~2.} Next, the user specifies the problem instance. Additional constraints can also be specified. The latter is interesting if the constraints defined by the problem definition should be modified. Problem-specific forms support the user in specifying the constraints.

\noindent
\textit{Step~3.} From the information in the previous step, this step estimates the number of qubits. For this purpose, the sum value of the number of binary variables is first determined. The basis for the calculation is a known QUBO formulation, as well as the specified problem instance. The sum value can in turn be used to estimate the number of (physically) required qubits. We store the calculated sum value in \textit{numVar}, the number of qubits in \textit{numQubits}. Both values are used in the next step.

\noindent
\textit{Step~4.} In this step, the appropriate solvers are selected from a set of stored solvers. Their technical data are matched with \textit{numVar} and \textit{numQubits}. QPU solvers have an upper limit in terms of qubits (attribute \textit{maxQubits} in Figure~\ref{fig:wfl}). If \textit{numQubits} is larger than the number of actually available qubits, then this QPU solver is not a candidate to solve this problem. Hybrid solvers are limited in the number of variables\footnote{Default values for \textit{maxVar}: \url{https://www.dwavesys.com/media/soxph512/hybrid-solvers-for-quadratic-optimization.pdf}} (attribute \textit{maxVar} in Figure~\ref{fig:wfl}). This value must be higher than \textit{numVar} for a hybrid solver to be considered.

\noindent 
\textit{Step~5.} In this step, the list of suitable solvers from Step~4 is sorted. The most promising solvers appear first in the sorting. The underlying sorting principle is as follows: If benchmarks exist for certain combinations of problem instances and solvers, the first step is to sort them in descending order according to the quality of the solution. Benchmark scores can be used to determine the quality of the solution. If no benchmarks exist, the results are sorted by solver types and quantum annealers with the highest number of qubits and the most interconnected hardware topology. 
Assuming that appropriate benchmarks exist, the first step is to sort the solvers using the one-dimensional Euclidean distance according to the following rule:

The following variables and definitions are given:

\begin{compactitem}
    \item $B = (B_{1},B_{2},...,B_{m}),\ m \in \mathbb{N}$: Represents the sequence of $m$ benchmark objects available for a specific optimization problem and refers to the problem sizes that were evaluated.
    \item Each benchmark object $B_{j}$ contains a fixed number $x$ of integer values that represent possible benchmark scores:
    \begin{equation}\label{eq_c5}
        B_{j} = (w_{j1},w_{j2},...,w_{jx}),\ j \in \{1,...,m \},\ x \in \mathbb{N}.
    \end{equation}
    Depending on the benchmark, $x$ can be variable and is the maximum number of tested solvers for a benchmark. To ensure a consistent data storage structure, the benchmarks are always sorted in ascending order by the value $w_{j1}$.
    \item $w_{j1} \in \mathbb{N}_{+}$: The value is called the 'main parameter', which represents the reference value being searched for and is fixed.~In relation to the benchmarks made for the traveling salesman problem, the value could be the number of nodes in a graph on which the test was carried out~\citep{Stogiannos2022}.
    \item $n \in \mathbb{N}_{+}$: Is the value entered by the user for which we are searching for the nearest $w_{j1}$. Assuming that we are looking specifically at the benchmarks for the traveling salesman problem, the value could be the number of nodes in a graph.
    \item A function that extracts the first value $w_{j1}$ from a benchmark object $B_{j}$: 
    \begin{equation}\label{eq_c6}
        f(B_{j})=w_{j1}, \forall j \in \{1,...,m\}.
    \end{equation}
    \item Set of all extracted values:
    \begin{equation}\label{eq_c7}
        M=(f(B_{j})~\vert~j =  1,...,m)=( w_{11},w_{21},...,w_{m1} ).
    \end{equation}
    This sequence only contains the first value of each benchmark object. As a restriction, benchmarks are currently only displayed for a single run.
    \item A function $c$ that defines an ordering relation that compares two elements $(x,y) \in M$ pairwise according to their distance from $n$.
    \begin{equation}\label{eq_c8}
    c(x,y)= 
        \begin{cases}
          -1, & \text{if $\vert x - n\vert < \vert y - n\vert$}\\
          0, & \text{if $\vert x - n\vert = \vert y - n\vert$}\\
          1, & \text{if $\vert x - n\vert > \vert y - n\vert$}\ 
        \end{cases}  
    \end{equation}
    where $x \neq y$ for each pair.
    \item A function $g: \mathbb{N} \rightarrow M$ with:
        \begin{equation}\label{eq_c9}
    g(c(x,y))= 
        \begin{cases}
          x, & \text{if $c(x,y) \leq 0$}\\
          y, & \text{if $c(x,y) > 0$}\ 
        \end{cases}  
    \end{equation}
     which returns an $x$ that is closest to $n$. If several values have the same distance from $n$, the first value $x$ found is used. If no $x$ is found, the first $y$ that does not deviate from $n$ by 10\% is used. Otherwise, the default sort order (as Step~2 below) will be used.
    \item $S = \{S_{1},S_{2},...,S_{u}\},\ u \in \mathbb{N}$: Represents the set of $u$ solver.
    \item Each solver $S_{j}$ is described by four characteristics that are relevant for the calculation: 
    \begin{compactitem}
        \item $s_{j} \in \mathbb{N}_{0}$: \textit{solutionQuality}
        \item $v_{j} \in \mathbb{N}_{0}$: \textit{maxVariables}
        \item $q_{j} \in \mathbb{N}_{0}$: \textit{maxQubits}
        \item $N_{j}$: Solver name.               
    \end{compactitem}
     There are other characteristics that are not initially relevant for the calculation.
\end{compactitem}

The benchmark with the appropriate value is then determined as follows:
\begin{equation}\label{eq_c10}
j^{*} = \min\{ j\ \vert\ j \in \{1,...,m \},\ w_{j1} = x~\text{or}~y~\text{(see Eq.~\ref{eq_c9})}\},
\end{equation}
where $j^{*}$ is the direct identification of the corresponding index of the benchmark. The benchmark object $B^{*}$ is then determined as follows:
\begin{equation}\label{eq_c11}
B^{*} = B_{j^{*}} = (w_{j^{*}1},w_{j^{*}2},...,w_{j^{*}x}).
\end{equation}
The \textit{solutionQuality} $s_{j}$ (i.e., the corresponding benchmark score, cf. Eq.~\ref{eq_c5}) is chosen from $B^{*}$ according to an assignment rule:
\begin{equation}\label{eq_c12}
s_{j}= 
    \begin{cases}
      w_{j^{*}2}, & \text{if $N_{j}=\text{''Solver 1''}$}\\
      w_{j^{*}3}, & \text{if $N_{j}=\text{''Solver 2''}$}\\
      w_{j^{*}4}, & \text{if $N_{j}=\text{''Solver 3''}$}\\
      w_{j^{*}5}, & \text{if $N_{j}=\text{''Solver 4''}$}\\
      \vdots & \text{(if there are other names, they will be assigned accordingly)}
    \end{cases}  
\end{equation}
The set of solvers $S^{*}$ after assigning the quality of the solution from all solvers is:
\begin{equation}\label{eq_c13}
S^{*}=\{ (S_{j},s_{j})~\vert~S_{j} \in S, s_{j}\ \text{according to the above rule} \}
\end{equation}
The sorted set is obtained by lexicographic sorting of $S^{*}$ using the following keys:
\begin{equation}\label{eq_c14}
\begin{split}
S_{i}^{*} \prec S_{j}^{*} \iff (s_{i} < s_{j})\ &\text{or}\ (s_{i} = s_{j}\ \text{and}\ v_{i} < v_{j})\ \\ &\text{or}\ (s_{i} = s_{j}\ \text{and}\ v_{i} = v_{j}\ \text{and}\ q_{i} < q_{j})
\end{split}
\end{equation}
The sorted set $S^{**}$ is then given by:
\begin{equation}\label{eq_c15}
S^{**} = (S^{*},\prec).
\end{equation}
As a second step, if there are no benchmarks, the solvers are sorted according to the following rule. Let the definitions of solvers $S$, \textit{maxVariables} $v_{j}$ and \textit{maxQubits} $q_{j}$ remain unchanged. Then the sorted set is obtained by the following hierarchical sorting:
\begin{equation}\label{eq_c16}
S_{i} \prec S_{j} \iff (v_{i} < v_{j})\ \text{or}\ (v_{i} = v_{j}\ \text{and}\ q_{i} < q_{j}).
\end{equation}
The sorted set $S^{*}$ is then given by:
\begin{equation}\label{eq_c17}
S^{*} = (S,\prec).
\end{equation}
\noindent
\textit{Step 6.} In the last step, it is the user's turn again. The user is informed of the estimated number of qubits required and receives a sorted list of solvers. More information on the solvers (such as prices) is given if available.
\section{Implementation and Evaluation Method}\label{prototype}
\begin{figure}[H]
 \centering
 \includegraphics[width=1.0\textwidth]{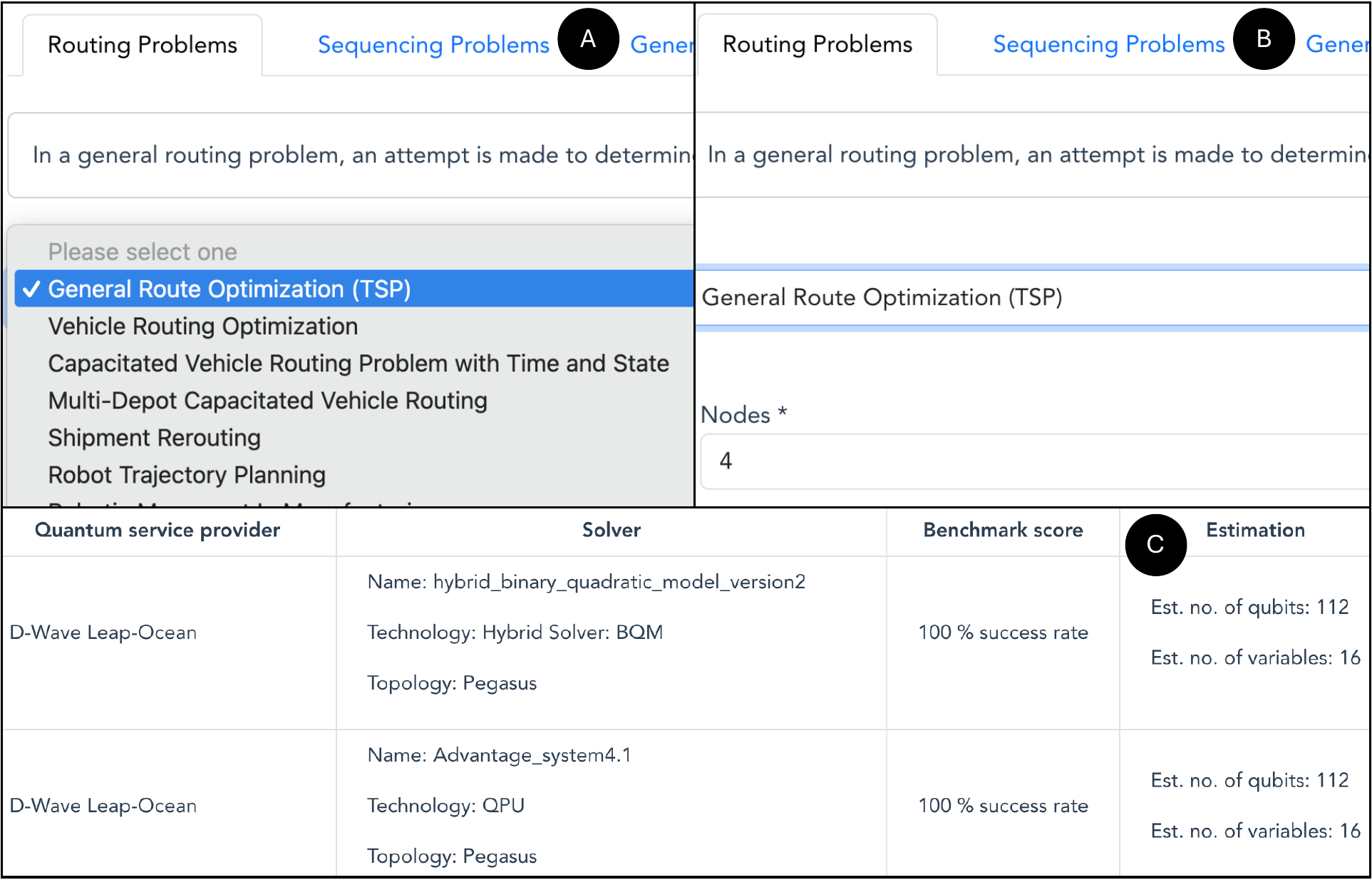}
 \caption{Compilation of the QC-Adviser user interface sections for workflow step 1 (Subsection~A:~Problem selection), workflow step 2 (Subsection~B:~Specification of the problem instance), and workflow step 6 (Subsection~C:~List of solvers found).}
 \label{fig:ui}
\end{figure}

In this section, we demonstrate how the prototype works using the TSP and describe the evaluation method. In doing so, we will go through the individual steps shown in~Figure~\ref{fig:wfl}. Figures~\ref{fig:ui}-A and ~\ref{fig:ui}-B show the QC-Adviser dialogs for user interactions.

\noindent
\textit{Step~1.} A user is interested in solving a TSP with a quantum annealer and would like a recommendation in this regard. To do so, the user selects TSP (cf. Figure~\ref{fig:ui}-A).

\noindent
\textit{Step~2.} The user specifies the TSP problem instance. The user is interested in the basic form of the problem, which corresponds to a complete graph (each pair of nodes is connected by an edge). Therefore, it is sufficient to specify only the number of nodes. The user decides for $4$ nodes (cf. Figure~\ref{fig:ui}-B). Note: Further development of the prototype will allow for the formulation of more complex TSP problems, including those that involve distance specifications, such as in the work of~Stogiannos et al.~\cite{Stogiannos2022}.

\noindent
\textit{Steps 3-5.} In these steps, the internal calculation of the QC-Adviser takes place without user interaction. The basis for computing \textit{numVar} and \textit{numQubits} for the TSP is based on the QUBO formulation mentioned in Section~\ref{preliminaries}. From the QUBO formulation, the formula $n^2$ for calculating \textit{numVar} was derived~\citep{Stogiannos2022}, where $n$ is the number of nodes specified in Step~2. Based on this, the number of qubits required is estimated, and the corresponding solvers are selected and sorted.

\noindent
\textit{Step 6.} The solvers are presented to the user in tabular form (cf. Figure~\ref{fig:ui}-C). The solver information, the estimated number of qubits required, and the benchmark score are displayed. The latter is only displayed if a benchmark score is available for the combination of problem instance and solver. The value indicates how often this solver correctly calculated the expected result (for a given number of runs). The user also has the option of calling up the usage prices.

\noindent
\textit{Some implementation aspects.} The prototype offers support for various problems (currently 15) and is not tailored to the TSP. The different QUBO formulations and the estimation approaches implemented come from the research literature. As examples, we mention the following research literature: For the TSP and the Vehicle Routing Problem the work of~\cite{Stogiannos2022} and~\cite{DBLP:conf/iccS/BorowskiGKBKMBS20}, for Job Shop Scheduling the work of~\cite{Carugno2022} as well as the work of~\cite{Lucas2014}, which contains a good overview of different problems. 

The prototype also has a modular structure and follows the MVC paradigm. This means that developers can add new problem classes, problems, and individual hardware datasets (for solvers) without affecting the overall application. The datasets are inserted in the form of JSON documents and form the \textit{QC-Services collection} in Figure~\ref{fig:wfl}. Fast-changing information, such as prices, can be obtained automatically via the service provider interface, if available.

\noindent
\textit{Evaluation method.} We outline the evaluation method using a problem for which a benchmark is available. If there are no benchmarks, the sorting is performed according to Eq.~\ref{eq_c16}.
\begin{compactenum}
    \item We define a problem instance (including conditions) and determine its best possible solution using classical methods. Such a solution is either known, obvious, or can be calculated in a reasonable amount of time if the problem instance is not too large. 
    \item We enter this problem instance into the QC-Adviser to obtain the sorted solvers. 
    \item Solution calculations for the given problem instance with at least two different solvers: To do this, we select two solvers. The first solver is at the top of the list, the second behind it, or at the end of the list. We calculate the solution to the problem instance using both solvers.
    \item Comparison of the best possible classical solution with the solutions determined by the solvers in order to evaluate the QC-Adviser (i.e., to ensure the correctness of the recommended solvers): Assume that the quality of the solution (benchmark score) of the first solver is 100\%. Then its solution should correspond to the classical best result (otherwise this solver would be incorrectly sorted because the top solvers deliver the best solutions).  And assuming that the quality of the solution of the second solver is significantly below 100\%. Then the result of this solver must deviate from the classically determined best result by a certain percentage (the further down a solver is in the list, the greater the deviation from the optimal result). In the case of TSP, this would mean that the determined tour takes longer than the optimally determined tour (i.e., a percentage deviation can be detected).
    \item In other problems, a different metric must be used, such as the makespan in the job shop scheduling problem. The overall evaluation is based on a systematic comparison of a classical problem solution with the solutions provided by the solvers for each problem, as well as the positions of the solvers.
\end{compactenum}
\section{Conclusion}\label{conclusion}
We have introduced the QC-Adviser prototype, a dialogue-based tool that supports users in selecting suitable solvers. Users only make domain-specific input and do not need any quantum computing know-how. The QC-Adviser results view lists the recommended solvers with basic information in a sorted list. The evaluation to date has been literature-based and selective. For a given problem instance and solution quality (benchmark scores) from at least two different solvers, we can check the expected positions of these solvers in the QC-Adviser list, as the solvers are sorted in descending order according to solution quality. The QC-Adviser tool is publicly available at Zenodo\footnote{\url{https://zenodo.org/records/17104675}}.

A comprehensive evaluation using the method described in Section~\ref{prototype} (with experiments conducted on real quantum hardware) will be considered in future work. In addition, the prototype can be further developed. More detailed problem-specific forms could be developed for the specification of the constraints. Additional solvers such as D-Wave's \textit{nonlinear solver} may also be considered.


\bibliographystyle{unsrt}  
\bibliography{references}  

\end{document}